\documentclass[aps,preprint,showpacs]{revtex4-1}
\usepackage{amsfonts}
\usepackage{amsmath}
\usepackage{amssymb}
\usepackage{graphicx}
\usepackage{rotating}
\usepackage{hyperref}
\usepackage{color}   %May be necessary if you want to color links
\usepackage[dvipsnames]{xcolor}

\begin{document}

\title{Magnetized color superconducting cold quark matter within the SU(2)$_f$ NJL model:
a novel regularization scheme}

\author{P. Allen$^{a}$, A.G. Grunfeld$^{a,b}$ and N.N. Scoccola$^{a,b,c}$ }
\affiliation{
$^a$ Department of Theoretical Physics, Comisi\'on Nacional de Energ\'ia
At\'omica, Av.Libertador 8250, 1429 Buenos Aires, Argentina\\
$^b$ CONICET, Rivadavia 1917, 1033 Buenos Aires, Argentina\\
$^c$ Universidad Favaloro, Sol\'is 453, 1078 Buenos Aires, Argentina}

\pacs{24.10.Jv, 25.75.Nq}

\begin{abstract}
The influence of intense magnetic fields on the behavior of color
superconducting cold quark matter is investigated using an
SU(2)$_f$ NJL-type model for which a novel regulation scheme is
introduced. In such a scheme the contributions which are
explicitly dependent on the magnetic field turn out to be finite
and, thus, do not require to be regularized. As a result of this,
non-physical oscillations that might arise in the alternative
regularization schemes previously used in the literature are
naturally removed. In this way, a clearer interpretation of the
physical oscillations is possible. The sensitivity of our results
to the model parametrization is analyzed.
\end{abstract}

\maketitle

\section{Introduction}

At asymptotically large chemical potentials, the fact that cold
quark matter behaves as a color superconductor can be shown by
using perturbative methods in the context of quantum
chromodynamics (QCD)\cite{Bailin:1983bm}. However, such methods
cannot be applied in the range of moderate densities relevant for,
amongst others, the astrophysics of strongly magnetized compact
stellar objects known as magnetars. We recall here that, although
it is generally accepted that these objects can have surface
magnetic fields up to $10^{15}$ G\cite{Duncan:1992hi}, the
estimates for the magnetic field values at their centers are model
dependent to some extent, ranging between $B \simeq 10^{18} -
10^{20}$ G (see e.g.
Refs.\cite{Shapiro:1991hi,Bandyopadhyay:1997kh,Ferrer:2010wz}).
Since the well-known sign problem prevents lattice QCD
calculations from being performed at sufficiently low temperatures
and finite chemical potential, one has to rely on effective models
to analyze the behavior of magnetized quark matter in this region.
One particular model that has been extensively used for this
purpose is the Nambu-Jona-Lasinio (NJL) model\cite{reports}. This
is an effective model originally devised to study the dynamics of
chiral symmetry breaking, in which gluon degrees of freedom are
integrated out and interactions are described by local four-quark
interactions. The incorporation of additional diquark interactions
into the model allows for the description of color superconducting
matter\cite{Buballa:2003qv}. In this context, the effect of a
constant magnetic field has been analyzed by several authors
\cite{Ferrer:2005vd,Fukushima:2007fc,Noronha:2007wg,Fayazbakhsh:2010gc,Fayazbakhsh:2010bh,Mandal:2012fq}.
At this point, it is important to remark that the local character
of the interactions considered in the NJL-type models leads to
divergences in the momentum integrals which need to be handled in
some way in order to completely define the model and yield
meaningful quantities. Several regularization procedures are
possible even in the absence of magnetic fields \cite{reports}.
Moreover, when magnetic field is introduced, the vacuum energy
acquires a Landau level (LL) structure and an additional care is
required in the treatment of the divergences. An elegant way of
treating the regularization has been reported in
Ref.\cite{Menezes:2008qt} for the model in the absence of color
superconductivity. The procedure follows the steps of the
dimensional regularization prescription of QCD, performing a sum
over all Landau levels in the vacuum term. This allows to isolate
the divergence into a term that has the form of the zero magnetic
field vacuum energy and that can be regularized in the standard
fashion. It should be stressed that similar expressions for the
magnetic field dependent terms can be obtained using a method
based on the proper-time formulation\cite{Ebert:1999ht}. So far,
however, this procedure has not been applied to the case in which
color superconductivity is present. Instead, existing calculations
\cite{Fukushima:2007fc,Noronha:2007wg,Fayazbakhsh:2010gc,Fayazbakhsh:2010bh,Mandal:2012fq}
remove the divergences by introducing some type of regulator
function for each Landau level separately. This procedure,
however, might in general introduce unphysical oscillations. A
discussion on this can be found in
Refs.\cite{Campanelli:2009sc,Frasca:2011zn,Gatto:2012sp}, where it
is also observed that the use of smooth regulator functions
improve the situation. In fact, this allows to identify possible
physical oscillations appearing in some
cases\cite{Fukushima:2007fc,Noronha:2007wg}. However, an even
clearer interpretation of the results could be obtained if the
unphysical oscillations were removed altogether with another
scheme, especially at finite chemical potential and in the
presence of color superconductivity. The main purpose of this work
is to investigate the influence of a constant magnetic field on
cold superconducting quark matter in the framework of the NJL-type
model, using a regularization procedure in which the contributions
that are explicitly dependent on the magnetic field turn out to be
finite and, thus, do not required to be regularized. This
procedure will be referred to as ``Magnetic Field Independent
Regularization" (MFIR), and it can be considered an extension of
the method described in e.g. Ref.\cite{Menezes:2008qt} to the case
in which color pairing interactions are present. Since the
aforementioned unphysical oscillations are completely removed in
this scheme, we will complement our analysis by performing a
detailed study of the resulting cold matter phase diagrams,
including their dependence on the parameters of the model and, in
particular, the coupling strength of the diquark interactions.
Being mostly concerned with the issues related to the model
regularization procedure we will, for simplicity, assume all the
quark species to have a common chemical potential leaving the
incorporation of the neutrality and $\beta$ equilibrium conditions
relevant for stellar matter applications for future studies.

We organize the article as follows. In Section II we present the
Nambu Jona Lasinio model with magnetic field and diquark
interactions. In particular, we  briefly review the regularization
schemes used in the literature and describe in some detail the
MFIR scheme introduced in this work. The model parameters used in
our numerical calculations are also given. In Sect. III we compare
our results for the behavior of the cold and dense magnetized
quark matter with those  previously reported in the literature. In
Sect. IV we present results for the phase diagrams in the $\tilde
eB-\mu$ plane as obtained using MFIR for different interaction
coupling ratios and parameter sets. In Sect. V we present our
conclusions. Finally, in the Appendix, several details of the
formalism of the MFIR scheme are described.

\section{Magnetized cold quark matter
within the SU(2)$_f$ NJL model in the presence of color pairing
interactions}

\subsection{The thermodynamical potential in the mean field approximation}

We consider a NJL-type SU(2)$_f$ Lagrangian density which includes
scalar-pseudoscalar and color pairing interactions. In the
presence of an external magnetic field and chemical potential it
reads:
\begin{eqnarray}
\mathcal{L} &=& \bar{\psi}\left[ i\ \tilde {\rlap/\!D}   - m_c + \mu \
\gamma^0 \right]\psi
\nonumber \\
& & \qquad  \qquad + G \left[ \left( \bar{\psi}\psi \right)^{2}+
\left( \bar{\psi}i\gamma_{5}\vec\tau\psi \right)^{2} \right] + H
\left[ (i \bar\psi^C \ \epsilon_f \epsilon^3_c \gamma_5 \psi) (i
\bar\psi \ \epsilon_f \epsilon^3_c \gamma_5 \psi^C) \right].
\label{equno}
\end{eqnarray}
Here, $G$ and $H$ are coupling constants,
$\psi=\left(u,d\right)^{T}$ represents a quark field with two
flavors, $\psi^C= C\bar\psi^T$ and $\bar \psi^C= \psi^T C$, with
$C=i\gamma^2\gamma^0$, are charge-conjugate spinors and $\vec
\tau= (\tau_1,\tau_2,\tau_3)$ are Pauli matrices. Moreover,
$(\epsilon_c^3)^{ab} = (\epsilon_c)^{3ab}$ and $(\epsilon_f)^{ij}$
are antisymmetric matrices in color and flavor space respectively.
Furthermore, $m_c$ is the (current) quark mass that we take to be
the same for both flavors and $\mu$ is the quark chemical
potential. The coupling of the quarks to the electromagnetic field
$\tilde {\cal A}_\mu$ is implemented through the covariant
derivative $\tilde D_{\mu}=\partial_\mu - i \tilde e \tilde Q
\tilde{\cal A}_{\mu}$. Note that here we are dealing with
``rotated" fields. In fact, as is well known, in the presence of a
non-vanishing superconducting gap $\Delta$, the photon acquires a
finite mass. However, as shown in Ref.\cite{Alford:1999pb}, there
is a linear combination of the photon and the eighth component of
the gluon field that leads to a massless rotated $U(1)$ field. The
associated rotated charge matrix $\tilde Q$ is given by
\begin{equation}
\tilde{Q} = Q_f \otimes 1_c - 1_f \otimes \left(\frac{\lambda_8}{2
\sqrt{3}}  \right)
\end{equation}
where $Q_f = \mbox{diag}(2/3,-1/3)$ and $\lambda^8$ is the color
quark matrix $\lambda_8 = \mbox{diag}(1,1,-2)/\sqrt{3}$. Then, in
a six dimensional flavor-color representation $(u_r, u_g, u_b,
d_r, d_g, d_b)$, the rotated $\tilde{q}$ for different quarks are:
$u_r=1/2, u_g=1/2, u_b=1, d_r=-1/2, d_g=-1/2, d_b=0$. The rotated
unit charge $\tilde e$ is given by $\tilde e = e \cos \theta$,
where $\theta$ is the mixing angle which is estimated to be
$\simeq 1/20$\cite{Gorbar:2000ms}. In the present work we consider
a static and constant magnetic field in the 3-direction, $\tilde
{\cal A}_\mu=\delta_{\mu 2} x_1 B$, which in fact is also a
mixture of the electromagnetic field and color fields.

In what follows we work in the mean field approximation (MFA),
assuming that the only non-vanishing expectation values are $<\bar
\psi \psi> = -(M - m_c)/2G$ and $<i \bar\psi^C \ \epsilon_f
\epsilon^3_c \gamma_5 \psi>=-\Delta/2H$, which can be chosen to be
real. Here, $M$ and $\Delta$ are the so-called dressed quark mass
and superconducting gap, respectively. The resulting MFA
thermodynamic potential at vanishing temperature reads
\begin{equation}
\Omega_{\mbox{\scriptsize{MFA} } } = \frac{(M-m_c)^2}{4G} +
\frac{\Delta^2}{4H} - \sum_{\left|\tilde{q}\right| =
0,\frac{1}{2},1} P_{\left|\tilde{q}\right|} \label{uno}
\end{equation}
where
\begin{eqnarray}
P_{\left|\tilde{q}\right| = 0} &=& \int \frac{d^3p}{(2 \pi)^3} \left[E_0^+ + \left|E_0^-\right|  \right],
\label{omega0}
\\
P_{\left|\tilde{q}\right| = 1} &=& \frac{\tilde eB}{8 \pi^2}
\sum_{k=0}^{\infty} \alpha_k \int_{-\infty}^{\infty} dp_z
\left[E_1^+ + \left|E_1^-\right|  \right], \label{omega1}
\\
P_{\left|\tilde{q}\right| = 1/2} &=&  \frac{\tilde eB}{4 \pi^2}
\sum_{k=0}^{\infty} \alpha_k \int_{-\infty}^{\infty} dp_z \left[
E_{1/2}^+ +  E_{1/2}^- \right] \label{omega12}.
\end{eqnarray}
Here, we have introduced $\alpha_k =2 - \delta_{k0}$ and
\begin{eqnarray}
E_0^{\pm} &=&  \sqrt{p^2 + M^2} \pm \mu
\nonumber \\
E_1^{\pm} &=& \sqrt{p_z^2 + 2k \tilde eB + M^2} \pm \mu
\nonumber \\
E_{1/2}^{\pm} &=& \sqrt{\left[\sqrt{p_z^2 +  k \tilde eB + M^2} \ \pm \ \mu\right]^2 + \Delta^2}
\label{seis}.
\end{eqnarray}

Clearly, Eqs.(\ref{omega0}-\ref{omega12}) are divergent and, thus,
require to be regularized. Some alternative schemes to achieve
this will be discussed in the following subsection. Given the
corresponding regularized form
$\Omega_{\mbox{\scriptsize{MFA}}}^{reg}$, the associated gap
equations for $M$ and $\Delta$ then read
\begin{equation}
\frac{\partial \Omega_{\mbox{\scriptsize{MFA}}}^{reg}
}{\partial(M,\Delta)}=0 \ .
\end{equation}
For each value of $\mu$ and $\tilde eB$, several solutions of
these equations will generally exist, corresponding to different
possible phases, and the most stable solution is that associated
to the absolute minimum of the thermodynamic potential.

\subsection{Regularization schemes}

As already mentioned, the contributions from
Eqs.(\ref{omega0}-\ref{omega12}) need to be regularized. In
previous
studies\cite{Fukushima:2007fc,Noronha:2007wg,Fayazbakhsh:2010gc,Fayazbakhsh:2010bh,Mandal:2012fq}
this was accomplished by introducing some cutoff function
$h_\Lambda(q)$ in the corresponding integrands, with $q=p$ in the
case of Eq.(\ref{omega0}) and $q=\omega_{p_z,k} \equiv \sqrt{p_z^2
+ 2k |\tilde q| B}$ for Eqs.(\ref{omega1},\ref{omega12}). The
obvious and simplest choice would be to take
$h_\Lambda(q)=\Theta(\Lambda-q)$. In what follows, we will refer
to this regularization scheme as ``sharp function regularization"
(ShFR). In this case, the integral in Eq.(\ref{omega0}), which is
magnetic field independent, is cut off when $p=\Lambda$. The
contributions coming from Eqs.(\ref{omega1}-\ref{omega12}) include
a sum over Landau levels, and the integral in each of these is cut
off for the $p_z$ that satisfies $\Lambda=\sqrt{p_z^2+2 k |\tilde
q| B}$, that is, such that the momentum and magnetic field
contribution to the quark dispersion relation does not exceed the
value of the cut-off. This would seem like a natural way of
extending the 3D sharp cutoff zero magnetic field regularization
to the finite $\tilde eB$ case. However, the magnetic field
dependence of the prescription brings in strong unphysical
oscillations. To minimize the effects of this magnetic field
dependence, the aforementioned studies have been carried out
replacing the Heaviside function with a smooth regulator. For
definiteness in this work, we consider the function $h_\Lambda(q)=
1/(1+\exp[(q/\Lambda-1)/a])$. We have verified that other possible
choices lead to similar results. To choose the value of the
constant $a$ that determines the regulator smoothness one is
limited by the fact that a too steep function does not improve
over the ShFR results and that a too smooth function leads to
values of the quark condensate in absence of the magnetic field
which are quite above the phenomenological range. Here, we follow
Refs.\cite{Fayazbakhsh:2010gc,Fayazbakhsh:2010bh} and consider
$a=0.05$. In what follows we will refer to this regularization
scheme as ``smooth function regularization" (SmFR). The
similarities and differences in the results yielded by the two
regularization schemes introduced so far will be discussed in the
following section. There it will be noted that, although somewhat
suppressed, the undesired oscillations are still present in SmFR.
It is important to remark at this stage that
$P_{\left|\tilde{q}\right| = 0,1}$ can be rewritten in terms of a
vacuum and a matter contribution, of which only the first one is
divergent. Thus, there is certain ambiguity on whether the
regularization function has to be included in the matter term or
not. Having analyzed both possibilities, we verified that in our
case, this ambiguity amounts at most to small quantitative
differences. The results to be presented correspond to the case
where only the vacuum energy is regularized.

To fully get rid of the above mentioned regularization artifacts,
we introduce in what follows an alternative scheme in which the
contributions that are explicitly dependent on the magnetic field
turn out to be finite and thus do not need to be regularized. We
will refer to this regularization scheme as ``magnetic field
independent regularization" (MFIR). We start by considering the
$P_{\left|\tilde{q}\right| = 0}$ contribution. Since it is
independent of the magnetic field, it can be treated in the usual
way \cite{reports}. Introducing a sharp 3D cutoff we get

\begin{eqnarray}
P_{\left|\tilde{q}\right| = 0} &=& \frac{1}{\pi^2} \int_0^\Lambda d p\ p^2 \ \sqrt{p^2+M^2} \nonumber\\
&+& \frac{\Theta(\mu - M)}{\pi^2} \left[ \frac{\mu (\mu^2
-M^2)^{3/2}}{3} - \frac{(\mu^2 -M^2)^2}{8}
h\left(\frac{M}{\sqrt{\mu^2 -M^2}} \right)  \right],
\label{omegazero}
\end{eqnarray}
where $h(z) = (2+z^2) \sqrt{1+z^2} + z^4 \ln[z/\sqrt{1+z^2}]$.

In the case of  $P_{\left|\tilde{q}\right| = 1}$ we note that,
except for the specific value of the quark charge, the
corresponding expression coincides with that analyzed in
Ref.\cite{Menezes:2008qt} where no color pairing interactions were
considered. Following the steps discussed in that reference
we get
\begin{eqnarray}
P_{\left|\tilde{q}\right| = 1} &=& \frac{1}{\pi^2}
\int_0^\Lambda d p\ p^2 \ \sqrt{p^2+M^2} +
\frac{\tilde eB}{4 \pi^2} \sum_{k=0}^{k_{max}} \alpha_k
\left[ \mu \sqrt{\mu^2 - s_k^2} -  s_k^2 \ln \left( \frac{\mu + \sqrt{\mu^2 - s_k^2}}{s_k} \right) \right]
\nonumber \\
& &
+ \frac{(\tilde eB)^2}{2 \pi^2}
\left[ \xi'(-1,x) + \frac{x-x^2}{2} \ \ln x + \frac{x^2}{4}   \right]
\label{omegamag1}
\end{eqnarray}
where $x = M^2/ (2 \tilde eB) $, $k_{max} = \mbox{Floor}[(\mu^2 -
M^2)/(2 \tilde eB)]$  and $s_{k}= \sqrt{M^2 + 2 k \tilde eB}$. In
Eqs.(\ref{omegazero}-\ref{omegamag1}), the first term is a vacuum
contribution which does not explicitly depend on the magnetic
field and the second term is the matter contribution. The last
term in Eq.(\ref{omegamag1}) is the explicit magnetic field
contribution to the vacuum, which has been isolated into a finite
term.

The case of ${\left|\tilde{q}\right| = 1/2}$ is more involved.
However, as discussed in detail in the Appendix, it can be cast
into the form
\begin{eqnarray}
P_{\left| \tilde{q} \right| = 1/2} &=& \frac{2}{\pi^2}
\int_0^\Lambda d p\ p^2 \ \left( E_\Delta^+ +  E_\Delta^- \right) +
\frac{(\tilde eB)^2}{2 \pi^2} \left[\xi'(-1,y) + \frac{y-y^2}{2} \ln y +
\frac{y^2}{4} \right]
\nonumber \\
& & + \frac{(\tilde eB)^2}{2\pi^2} \int_0^\infty dp \ \left[ \sum_{k=0}^\infty \alpha_k \ f(p^2+ k) - 2 \int_0^\infty dx\ f(p^2+ x) \right]
\label{equa}
\end{eqnarray}
where $E_\Delta^\pm = \sqrt{(\sqrt{p^2+M^2}\pm \mu)^2+\Delta^2}$, $y = (M^2+\Delta^2)/ (\tilde eB) $ and
\begin{equation}
f(z) = \sum_{s=\pm1} \left[
\sqrt{ (\sqrt{z + 2 x} + s\ \mu/\sqrt{\tilde eB})^2 + y - 2 x} - \sqrt{z + y} \right].
\end{equation}
In this expression a 3D sharp cutoff has been introduced to
regularize the first term, i.e. the one that contains
contributions from the vacuum and matter which do not explicitly
depend on the magnetic field. Note that, as in the case of
vanishing $\tilde eB$ discussed in e.g.
Ref.\cite{Blaschke:2003cv}, these cannot be disentangled into two
terms unless $\Delta=0$. The second term is the vacuum magnetic
contribution analogous to the $|\tilde q|=1$ case. Finally, the
third term is an additional explicitly magnetic field dependent
matter contribution which, as shown in the Appendix, turns out to
be finite.

\subsection{Model parametrization}

In order to analyze the dependence of the results on the model
parameters, we will consider two SU(2)$_f$ NJL model
parameterizations. Set~1 corresponds to that leading to
$M_0=340$~MeV while set~2 to that leading to $M_0=400$~MeV, within
the MFIR regularization. Here, $M_0$ represents the vacuum quark
effective mass in the absence of external magnetic fields. The
corresponding model parameters are listed in Table~\ref{pnjl}.

\begin{table}[h]
\caption{\label{pnjl} Parameter sets for the SU(2)$_f$ NJL model.
In both cases, empirical values in vacuum for the pion observables
are reproduced, $m_{\pi}=138$MeV and $f_{\pi}=92.4$MeV. }
\begin{center}
\begin{tabular}{cccccc}
  \hline  \hline
\hspace*{.2cm} Parameter set \hspace*{.2cm} & \hspace*{.2cm} $M_0$
\hspace*{.2cm}& \hspace*{.2cm} $m_c$ \hspace*{.2cm} &
\hspace*{.2cm} $G\Lambda^2$ \hspace*{.2cm}    & \hspace*{.2cm}
$\Lambda$ \hspace*{.2cm}    &
\hspace*{.2cm} $-<u\bar u>^{1/3}$ \hspace*{.2cm}  \\
         &    MeV    &   MeV    &         &  MeV   &  MeV  \\ \hline
Set 1    &    340    &  5.595   & 2.212   &  620.9 & 244.3 \\
Set 2    &    400    &  5.833   & 2.440   &  587.9 & 240.9
\\
\hline
\end{tabular}
\end{center}
\end{table}

We should observe that in the $\tilde eB=0$ limit, ShFR and MFIR
regularization schemes result in the same mass. On the other hand,
the masses for SmFR are about $10$ MeV larger for both sets of
parameters. This happens because the regulator function has a
non-zero tail for large momentum with respect to the ShFR case,
which causes the contribution from the vacuum to be larger, thus
giving rise to a somewhat larger dressed mass.

\section{Numerical results}

\subsection{Comparison between regularization schemes}

To motivate the introduction of the MFIR scheme, in this
subsection we will compare the resulting predictions with the ones
of the ShFR and SmFR. For definiteness, we will consider set~1 and
take $H/G = 0.75$, value that follows from various effective
models of quark-quark interactions\cite{Buballa:2003qv}. In
addition, some general features of the results will be described.

\begin{figure*}
\begin{center}
\includegraphics[scale=0.5]{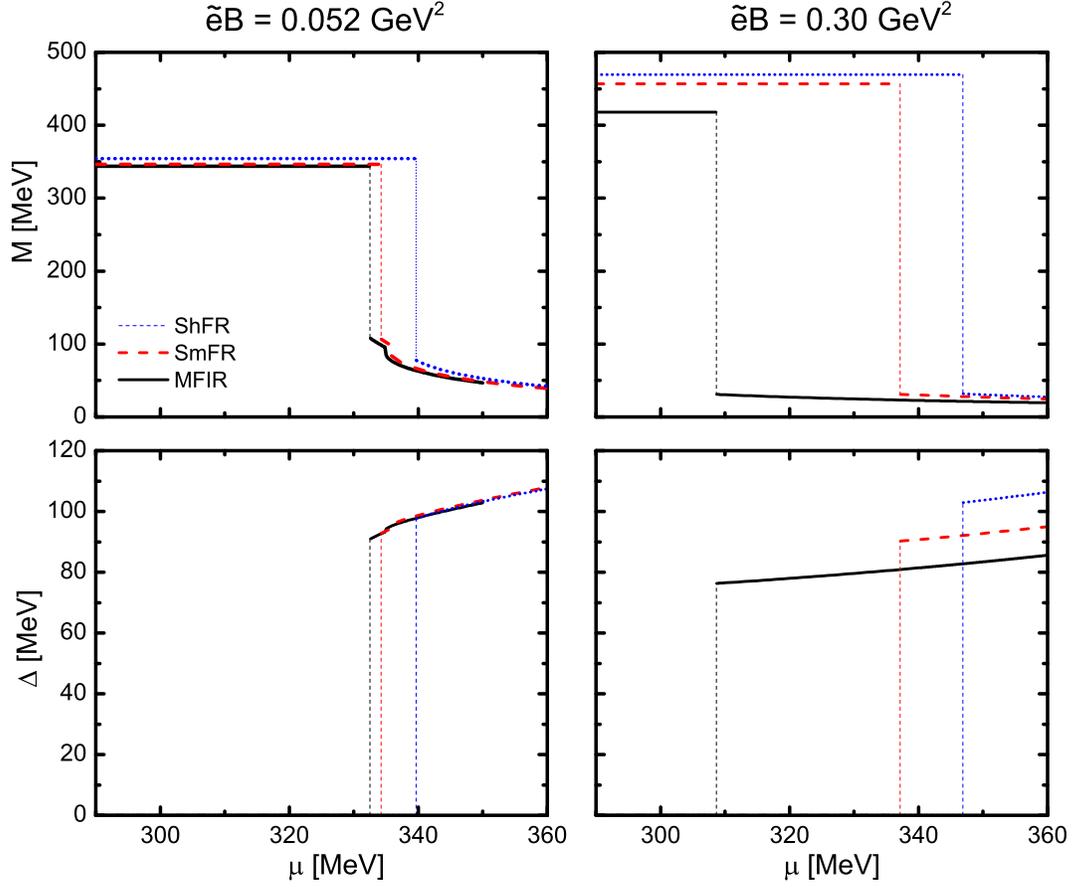}
\end{center}
\caption{ Left panels: $M$ vs $\mu$ (upper) and $\Delta$ vs $\mu$
(bottom), for $\tilde eB=0.052$~GeV$^2$. Right panels: $M$ vs
$\mu$ (upper) and $\Delta$ vs $\mu$ (bottom), for $\tilde
eB=0.30$~GeV$^2$. Set~1 was used, $H/G=0.75$ and all
regularization schemes considered.} \label{fig:1}
\end{figure*}

In Fig.~\ref{fig:1}, we plot the results for $M$ and $\Delta$ as a
function of $\mu$, for all three regularizations. At a fixed
magnetic field, these would appear to present similar behaviors.
Two distinct types of phases exist in all cases: on the one hand,
for low chemical potential, chiral symmetry is broken and
superconducting effects are absent. Since $M > \mu$, all matter
terms are zero and the dressed mass is independent of the chemical
potential. This phase, to be denoted as $\mbox{B}$ phase as in
previous
studies\cite{Ebert:1999ht,Ebert:2003yk,Allen:2013lda,Allen:2015qxa,Grunfeld:2014qfa},
always exists for a low enough chemical potential. If chemical
potential is increased, on the other hand, there will be a first
order phase transition (whose critical chemical potential $\mu_c$
depends on the regularization) to a phase where $\Delta$ is non
vanishing and the dressed mass is small. It is considered to be a
restored symmetry phase, even though exact symmetry restoration
occurs only if $m_c=0$. Since in our case the current quark mass
is finite but small, restoration is only approximate. For higher
chemical potentials, further transitions may appear within the
$\Delta \neq 0$ region, causing it to have a substructure
consisting of several phases. One of such transitions can be seen
in the upper left panel of Fig.~\ref{fig:1}, signalled by a small
kink in the mass at $\mu_c=334.9$~MeV. To understand the origin of
these we must recall that, for the quark species with $|\tilde
q|=1/2$ and $1$, the dispersion relations acquire a Landau level
structure due to the magnetic field, as seen in Eq.(\ref{seis}).
Let us consider first the case $|\tilde q|=1$, where in the matter
contribution to the thermodynamic potential a sum over these
Landau levels up to $k_{max}$ has to be performed. Here, $k_{max}$
is determined by the chemical potential, mass and magnetic field.
Namely, $k_{max} = \mbox{Floor}[(\mu^2 - M^2)/(2 \tilde eB)]$.
Following the notation of previous works, a chirally restored
phase where the LL's are populated up to a given $k$ will be
referred to as an $\mbox{A}_k$ phase, even though we bear in mind
that this phase is qualitatively different in that there is a
finite diquark gap now. The kink in Fig.~\ref{fig:1} then
corresponds to a transition in which the highest LL that is
populated (given by $k_{max}$) changes in one unit. In what
follows, we will refer to these as ``van Alphen-de Haas (vAdH)
transitions". It is also clarifying to mention that the dressed
mass vanishes in the chiral case, so the vAdH transitions are
actually signalled by discontinuities in the density, and that for
the $k^{th}$ LL they are simply given by the relation $\mu=\sqrt{2
k |\tilde q| B}$. In the non-chiral case there will be a small
departure from this relation originating from the finite mass in
the restored phase. It is important to stress that out of the
quark species with different values of $|\tilde q|$, the only one
that produces vAdH transitions is $|\tilde q|=1$. In fact,
$|\tilde q|=0$ quarks are decoupled from the magnetic field, so
their dispersion relation is the same as in the zero magnetic
field case. Moreover, quarks with $|\tilde q|=1/2$ have an
altogether different behavior. The coupling of this quark species
to the $\Delta$ removes the theta functions from the sum, much in
the same way that a theta function becomes a Fermi-Dirac
distribution when temperature is introduced. Hence, there is no
cut off in the sum over LL´s, which means that when $\Delta$ is
finite there is non-zero density for all levels.

\begin{figure*}
\begin{center}
\includegraphics[scale=0.55]{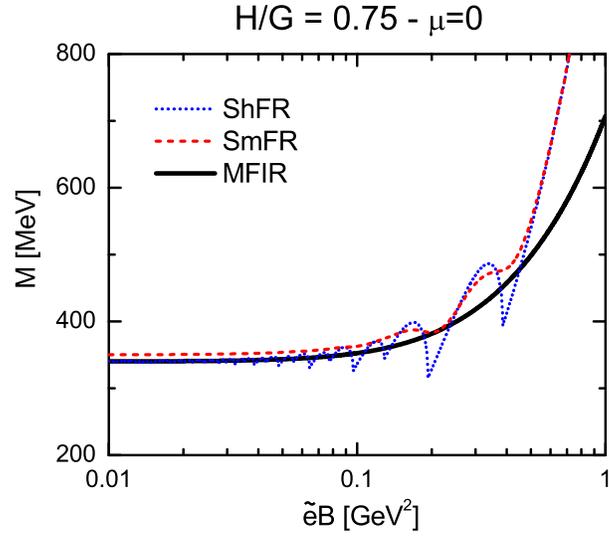}
\end{center}
\caption{$M$ vs $\tilde eB$ in the $\mbox{B}$ phase, for the three
regularization schemes considered, set~1 and $H/G=0.75$. Note that
$\Delta=0$ in this phase.} \label{fig:2}
\end{figure*}

\begin{figure*}
\begin{center}
\includegraphics[scale=0.5]{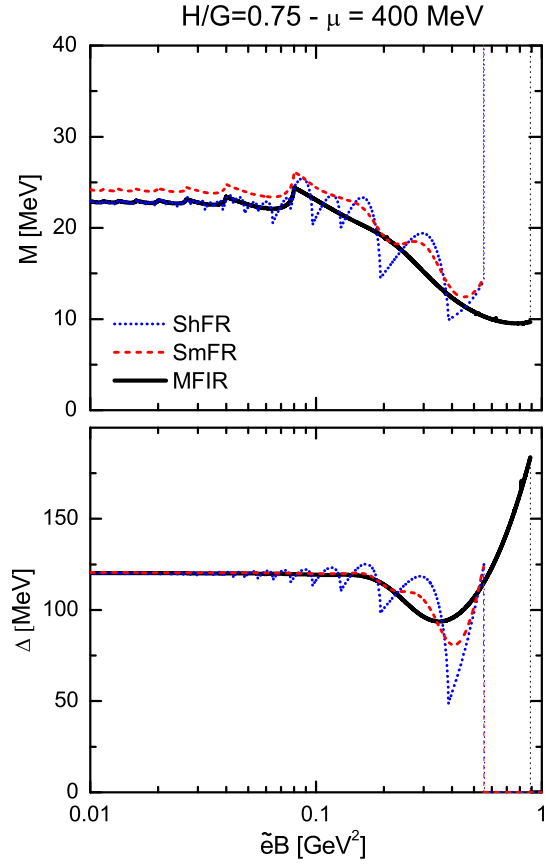}
\end{center}
\caption{$M$ and $\Delta$ vs $\tilde eB$ in the A-type region, for
the three regularization schemes considered, set~1 and
$H/G=0.75$.}
\label{fig:3}
\end{figure*}

Now, even though the order parameters have similar behaviors at a
fixed $\tilde eB$ as a function of chemical potential, the three
regularizations exhibit important qualitative differences along
the magnetic field axis, as can be seen in Figs.~\ref{fig:2}
and~\ref{fig:3}. Regularization schemes ShFR and SmFR exhibit
non-physical oscillations, whose origin lies in the magnetic field
dependence of the regularization in the vacuum term, that causes
the contribution of a given LL to be larger for lower magnetic
fields. This can be most clearly appreciated in Fig.~\ref{fig:2},
which displays the behavior of $M$ as a function of $\tilde eB$
for $\mu=0$. Here, the only contribution to the thermodynamic
potential comes from the regularized vacuum. In the ShFR, which is
the most extreme case, the only LLs participating in the sum are
those for which $\Lambda^2 \geq  2 k |\tilde q| B$. Hence,
depending on the magnetic field, more or less terms appear and
each time the relation is satisfied for a given $k$, there will be
a discontinuity in the derivative of the thermodynamic potential.
This singularity, hence, does not correspond to a phase
transition. The soft regulator, which could be regarded as a way
to handle this problem and remove sharp oscillations, still
contains this pathology, because in this case the contribution of
a given LL also depends on the magnetic field through the
Fermi-type regulator function. So, even though the smooth
integrals partly conceal this problem, the oscillations are still
present and in Fig.~\ref{fig:2} we can actually see that for ShFR
and SmFR they are in phase. On the other hand, in the MFIR scheme
the mass increases steadily with the magnetic field displaying the
usual ``magnetic catalysis effect" as in e.g.
Refs.\cite{Menezes:2008qt,Ebert:2003yk,Allen:2015qxa,Boomsma:2009yk}.

The behavior of the vAdH transitions can also be appreciated in
Fig.~\ref{fig:3}, where we plot $M$ and $\Delta$ as functions of
$\tilde eB$ for $\mu=400$~MeV. In the range $\tilde
eB=0.01-0.1$~GeV$^2$ there is a set of peaked discontinuities,
each of which corresponds to a vAdH transition. It should be
emphasized that these are physical transitions, as opposed to the
discontinuities previously discussed, since they correspond to
values of $\tilde eB$ at which the quark density for a given LL
changes from zero to a finite number. Note that in the ShFR scheme
they are harder to see because the non-physical oscillations
originating from the vacuum contribution are of the same order of
magnitude. In the MFIR scheme we can also observe that within each
phase, for a given finite $k$, the mass tends to decrease when
$\tilde eB$ increases, but it shows a small increase before the
next jump. In the $k=0$ phase, the mass decreases steadily with
magnetic field as well. In the chirally restored phase, the
superconducting gap will not vanish, and its behavior as a
function of magnetic field is nontrivial, as shown in
Fig.~\ref{fig:3}. Also, note that $\Delta$ is approximately
constant in the range of $\tilde eB \lesssim 0.12 \ \mbox{GeV}^2$,
with small oscillations resulting from its coupling to the mass,
and then presents a well-shaped curve.

\begin{figure*}
\begin{center}
\includegraphics[scale=0.55]{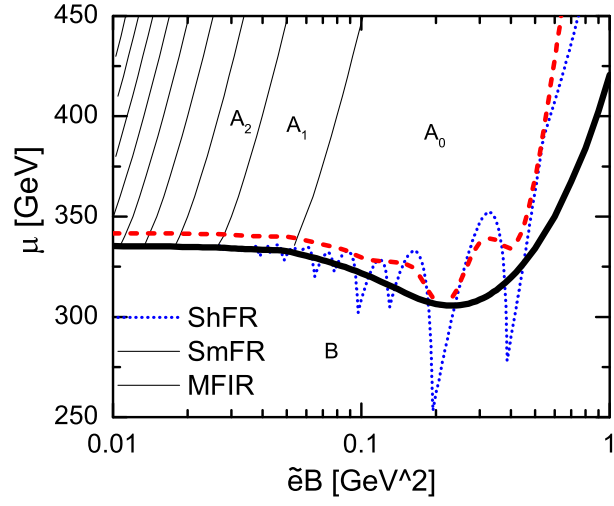}
\end{center}
\caption{Phase diagram in the $\tilde eB$ vs $\mu$ plane, for the
three regularization schemes considered, set~1 and $H/G=0.75$. All
transitions seen in this diagram are first order. } \label{fig:4}
\end{figure*}

The phase diagrams in the $\tilde e B-\mu$ plane as obtained using
the three different regularization schemes are displayed in
Fig.~\ref{fig:4}. In the case of the ShFR and SmFR, we see that
the oscillations in the order parameters induce oscillations in
the critical chemical potential. These are small for low magnetic
fields, but become larger in the intermediate $\tilde eB$ range
and once again make the phase diagram hard to interpret. As a
result, we can also conclude that comparing the three
regularizations for a given $\tilde eB$, as was done in
Fig.~\ref{fig:1}, is actually misleading, since the oscillating
behaviors in the order parameters and the critical chemical
potential in ShFR and SmFR can cause the results to look quite
different even for magnetic fields which are slightly different.
In the MFIR scheme, we note that the critical chemical potential
is approximately independent of $\tilde eB$ for values below
$0.07$~GeV$^2$, then it decreases until it reaches a minimum near
$\tilde eB=0.2 \ \mbox{GeV}^2$ and after this value, it increases
back again, giving rise to the usual well-shaped curve related to
the ``inverse magnetic catalysis effect"\cite{Preis:2010cq}. Due to the
regulation artifacts, this feature is much less evident in the
ShFR and SmFR. Concerning the vAdH transitions, which are the near
vertical lines, we note that they are almost equal for the three
prescriptions. In Fig.~\ref{fig:4} they actually correspond to the
MFIR case, but we make the observation that near the chiral
restoration transition small deviations exist, which occur because
the value of $M$ that produces the deviation with respect to
$\sqrt{2 k \tilde eB}$ is different in each scheme.

\subsection{MFIR results for different model parameters}

In this subsection we further analyze the results obtained within
the MFIR scheme, paying particular attention to their dependence
on the model parameters. Results for the two parametrizations
introduced in Sec.III.C will be given. Moreover, in the previous
section only $H/G = 0.75$ was considered. However, given that the
value of this ratio is subject to certain degree of uncertainty,
it is worthwhile to explore the consequences of varying it within
a reasonable range. Thus, in what follows, the representative
values $H/G= 0.5, 0.75$ and $1$ will be considered. A few comments
on how the model results change for $H/G < 0.5$ will be also made.
Note that values $H/G >1$ are quite unlikely to be realized in
QCD.

\begin{figure*}
\begin{center}
\includegraphics[scale=0.50]{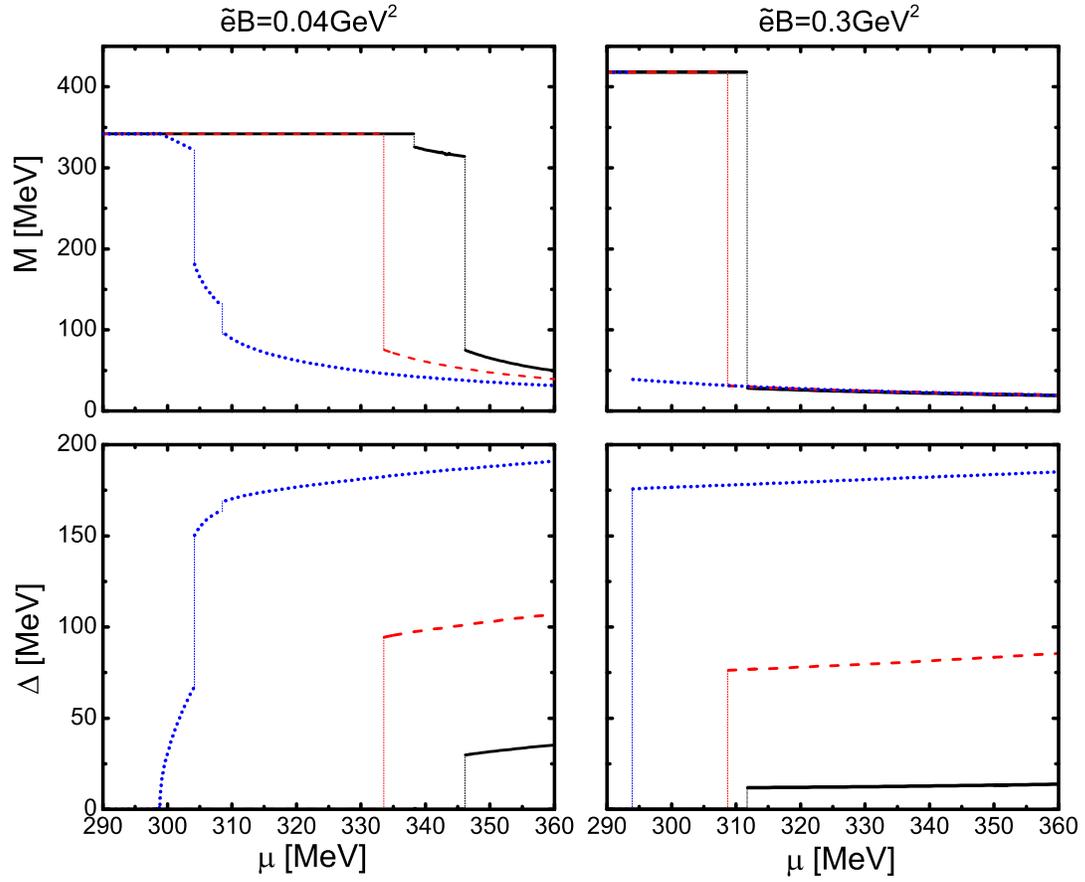}
\end{center}
\caption{M and $\Delta$ vs $\mu$ within the MFIR scheme for set~1,
different $H/G$ ratios and two values of $\tilde eB$. Full, dashed
and dotted lines correspond to $H/G = 0.5, 0.75$ and $1$
respectively.} \label{fig:5}
\end{figure*}

Let us start by analyzing the behavior of the order parameters as
a function of $\mu$ for given values of magnetic field. We will
concentrate on the results obtained with set 1, since it exhibits
a more complex phase structure. Set 2 will be addressed further
on. As was seen in the previous section for $H/G=0.75$, the system
is in the $\mbox{B}$ phase for low $\mu$, where the dressed mass
is large. On the other hand, it is in one of the possible
$\mbox{A}$-type phases for a high enough $\mu$ value where the
dressed mass is small. However, if the coupling ratio is changed,
other phases may appear for low $\tilde eB$ values and
intermediate chemical potentials. This can be seen in the left
panels of Fig.~\ref{fig:5}, where $\tilde eB = 0.04$~GeV$^2$. For
$H/G=0.5$, there is at $\mu_c=338.2$~MeV a weak first order
transition from vacuum to a phase where the mass is slightly lower
and also a slowly decreasing function of $\mu$. Quark density is
finite for the $|\tilde q|=0$ and $|\tilde q|=1$ quarks (only the
lowest LL being occupied for the latter species). Following
Refs.\cite{Ebert:2003yk, Ebert:1999ht,
Allen:2013lda,Allen:2015qxa,Grunfeld:2014qfa} this phase will be
denoted as a $\mbox{C}$-type phase although, as in the case of the
$\mbox{A}$-type phases, here the superconducting gap is non zero.
Actually, it happens to be very small, remaining always under
$1$~MeV. Hence, it is not visible in this scale. If $\mu$ is
further increased, we find another first order transition to an
A-type phase, at $346.1$~MeV. However, we note that $\Delta$ is in
the range $25-30$~MeV, which is a relatively small value compared
to the resulting ones from higher coupling ratios in
$\mbox{A}$-type phases. As the coupling constant ratio is
increased from the value $H/G=0.5$, the upper phase transition
displaces downwards. This causes the $\mbox{C}$-type phase to
shrink until it eventually disappears around $H/G\sim0.65$, so
that a single phase transition remains connecting the $\mbox{B}$
phase to the $\mbox{A}$-type phases. As the coupling ratio is
further increased, this phase transition continues to move
downwards and for $H/G=0.75$, the transition occurs at
$\mu_c=333.6$~MeV. For even larger values of $H/G$ ($\sim 0.94$),
the phase transition splits into two once again, so that for
$H/G=1$ there is an intermediate phase, which will be referred to
as a $\mbox{D}$ phase. This phase is qualitatively different from
the one found in $H/G=0.5$. To begin with, the transition from
vacuum to this phase is second order. The dressed mass is still
large but the superconducting gap is finite and actually increases
sharply with $\mu$. Since both condensates are appreciably large,
this is usually referred to as a ``mixed phase"
\cite{Huang:2001yw}, even though other meanings exist in the
literature for this term \cite{Neumann:2002jm}. There is no quark
population for the $|\tilde q|=0$ and $|\tilde q|=1$ species, but
the finiteness of $\Delta$ induces a non-zero density for $|\tilde
q|=1/2$ quarks. The transition leading to the $\mbox{A}$-type
phases is first order as in the previous cases and
$\Delta\sim175$~MeV. Within it, there is another transition for
$\mu=308.5$~MeV, which is actually a vAdH being traversed
vertically. The behavior is much simpler for $\tilde eB =
0.3$~GeV$^2$, as can be seen in the right panels of
Fig.~\ref{fig:5}. For all values of the coupling ratio, we only
see the $\mbox{B}$ and $\mbox{A}_0$ phases. Even though
transitions to higher $\mbox{A}_k$ phases will appear for much
higher chemical potentials, it is generally seen that the phase
structure is simpler for $\tilde eB \gtrsim 0.15$~GeV$^2$, so that
a single transition connecting the $\mbox{B}$ phase to the
$\mbox{A}_0$ phase is seen in the range of $\mu$ which is of
interest (see \cite{Allen:2013lda} for a detailed discussion on
how this occurs in NJL with magnetic field and without diquark
pairing). For both magnetic fields, we see that increasing $H/G$
always produces larger values for $\Delta$ and reduces the $\mu_c$
necessary to achieve the superconducting phase. We should also
note that for smaller magnetic fields, $\Delta$ grows faster as a
function of $\mu$ than for $\tilde eB=0.3$~GeV$^2$, where it seems
to be almost constant. As a matter of fact, we also checked that
for magnetic fields larger than a value around $0.4$~GeV$^2$, the
superconducting gap actually decreases (yet only slightly) with
chemical potential.

\begin{figure*}
\includegraphics[scale=0.50]{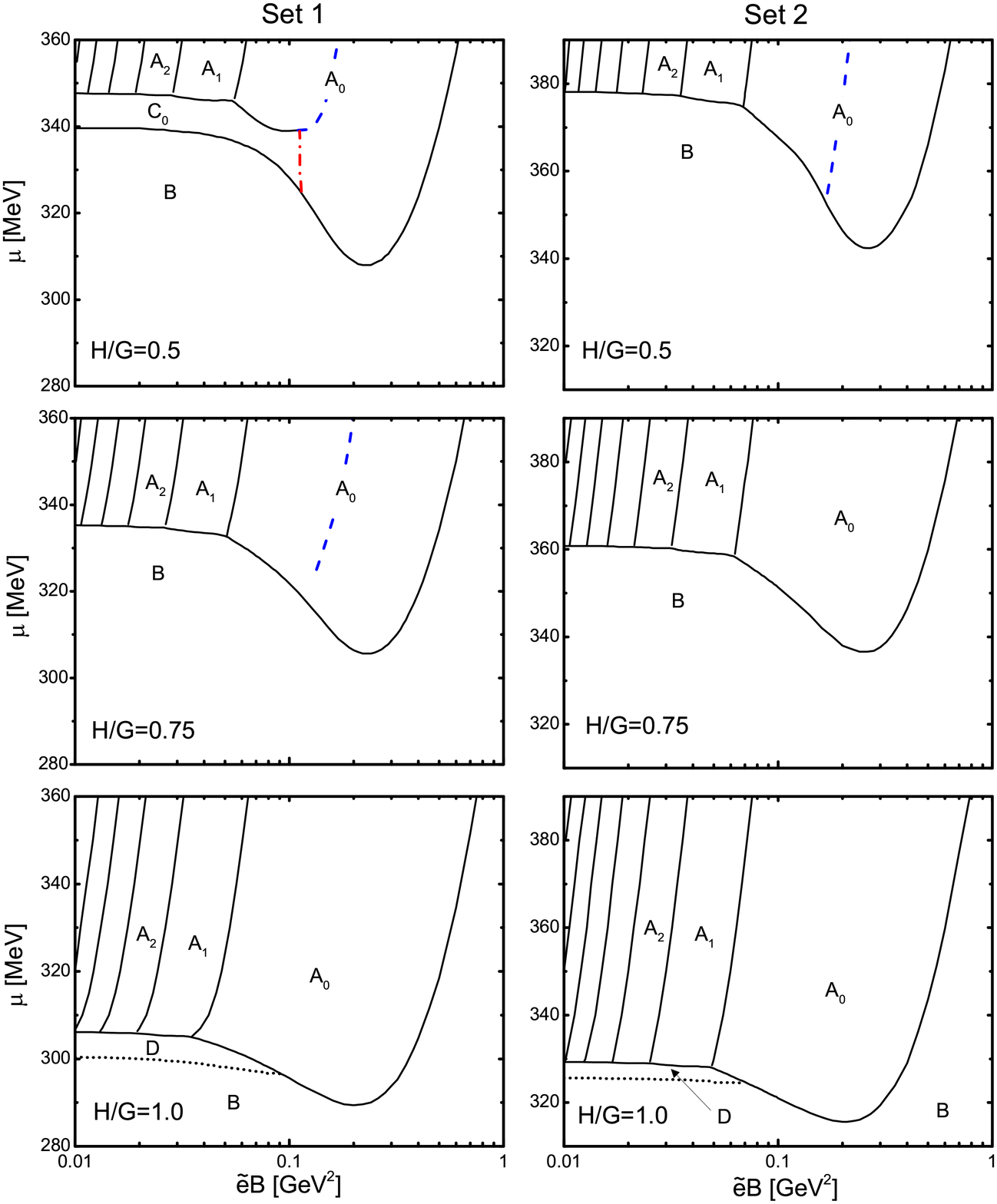}
\caption{Phase diagrams within the MFIR scheme for different $H/G$
ratios, for both set~1 (left panels) and set~2 (right panels).
Full black lines correspond to first order transitions and dotted
black lines to second order transitions. The dash-dotted red lines
represent chiral susceptibility crossovers. Blue dashed lines
represent quark number susceptibility crossovers which are too
weak to be actually considered to separate distinct phases. Note
that the sets are plotted in different intervals of the $\mu$
axis, even though the scale is the same.} \label{fig:6}
\end{figure*}

A deeper understanding of the behavior of the phases can be
obtained from the phase diagrams in the $\tilde eB - \mu$ plane
(Fig.~\ref{fig:6}), where our three coupling ratios and both
parameter sets are considered. The first transition encountered if
the phase diagram is traversed in the direction of increasing
$\mu$ will be referred to as the ``main transition". It has
approximately the same shape in all displayed phase diagrams and
it connects the $\mbox{B}$ phase to populated phases in general.
In set 2, the fact that $M_0$ adjusts to a higher value causes the
main transition to occur at a higher chemical potential. On the
other hand, as $H/G$ increases, the main transition is displaced
downwards in its entirety, and the depth of the ``inverse magnetic
catalysis well" diminishes. In set 1, we can see for $H/G=0.5$
(top left panel) that both transitions enclosing the
$\mbox{C}$-type phase are constant in a large magnetic field range
so that it extends in an approximately horizontal band up to
$\tilde eB \simeq0.11$~GeV$^2$, where it is bounded by a crossover
type transition, signalled by the peak of the chiral
susceptibility. The crossover leads to an $\mbox{A}_0$ phase and,
as expected, the mass drops sharply and $\Delta$ increases. The
different possible criteria to define this kind of transitions
were discussed in detail in \cite{Allen:2013lda}. On the other
hand, if magnetic field is decreased down to zero, the two
horizontal transitions continue to exist (however, the lower
transition becomes second order when $\tilde eB=0$), which means
that a phase with large mass and quark population exists in the
NJL model without magnetic field. The detailed behavior of the
$\mbox{C}$-type region is rather involved in the limiting case in
which $\tilde eB$ tends to zero, and will not be discussed any
further. It is important to point out that the existence of
$\mbox{C}$-type phases is not a consequence of the diquark pairing
channels either. As a matter of fact, up to $H/G=0.5$
superconducting effects are still small, and the phase diagram
remains basically unchanged. In fact, the two roughly horizontal
transition lines are seen to remain almost unmodified if the
coupling ratio is swept between these two parameters. We observe
that even though this $\mbox{C}$-type region consists of a single
phase where only the lowest LL is populated (hence it is a
$\mbox{C}_0$ phase), the results from
\cite{Allen:2013lda,Allen:2015qxa} suggest that it may become more
complex if vector interactions are introduced, or for parameter
sets leading to a lower $M_0$ value. In that case, it could be
possible that $\mbox{C}_k$ phases with higher $k$'s appear between
the two horizontal transitions. In the $\mbox{A}$-type phases,
higher LL population is allowed because the mass is lower than in
the $\mbox{C}$-type phases. vAdH transitions separate phases with
different number of LL populated, and in the presented diagrams,
LL's up to $k=7$ are occupied by $|\tilde q|=1$ quarks. On the
other hand, the phase diagram is simpler in set 2 (top right
panel), where it is seen that there is no intermediate phase for
$H/G=0.5$. In \cite{Allen:2013lda} we showed that changing the
parameter set so that $M_0$ increases always produces a simpler
phase structure where there are only a main transition and vAdH
transitions. As has been said, the upper transition found in set 1
moves downwards when $H/G$ is increased, so the $\mbox{C}_0$ phase
shrinks and finally disappears, leaving a rather simple phase
structure for $H/G=0.75$ (middle-left) panel. Note that for this
value of the coupling ratio the phase diagrams for both sets are
qualitatively similar. Finally, the phase diagrams for $H/G=1$ are
shown in the bottom panels, where the $\mbox{D}$ phase is seen to
exist for both parameter sets. Once again, we observe that the
existence of this phase is a consequence of the diquark pairing
alone and hence it already exists for $\tilde eB=0$. It is
interesting to note, however, that it continues to exist for
finite magnetic fields. The two transitions delimiting it are
roughly horizontal for low $\tilde eB$. For higher values, the
transitions lines move closer together and finally intersect near
$0.1$~GeV$^2$ causing the D phase to disappear. At low magnetic
field, the $\mbox{D}$ phase exists for a narrow $\mu$ range of at
most $5$~MeV for either set, but larger values of $H/G$ would
cause the phase to extend farther in the chemical potential
direction.

\begin{figure*}
\begin{center}
\includegraphics[scale=0.50]{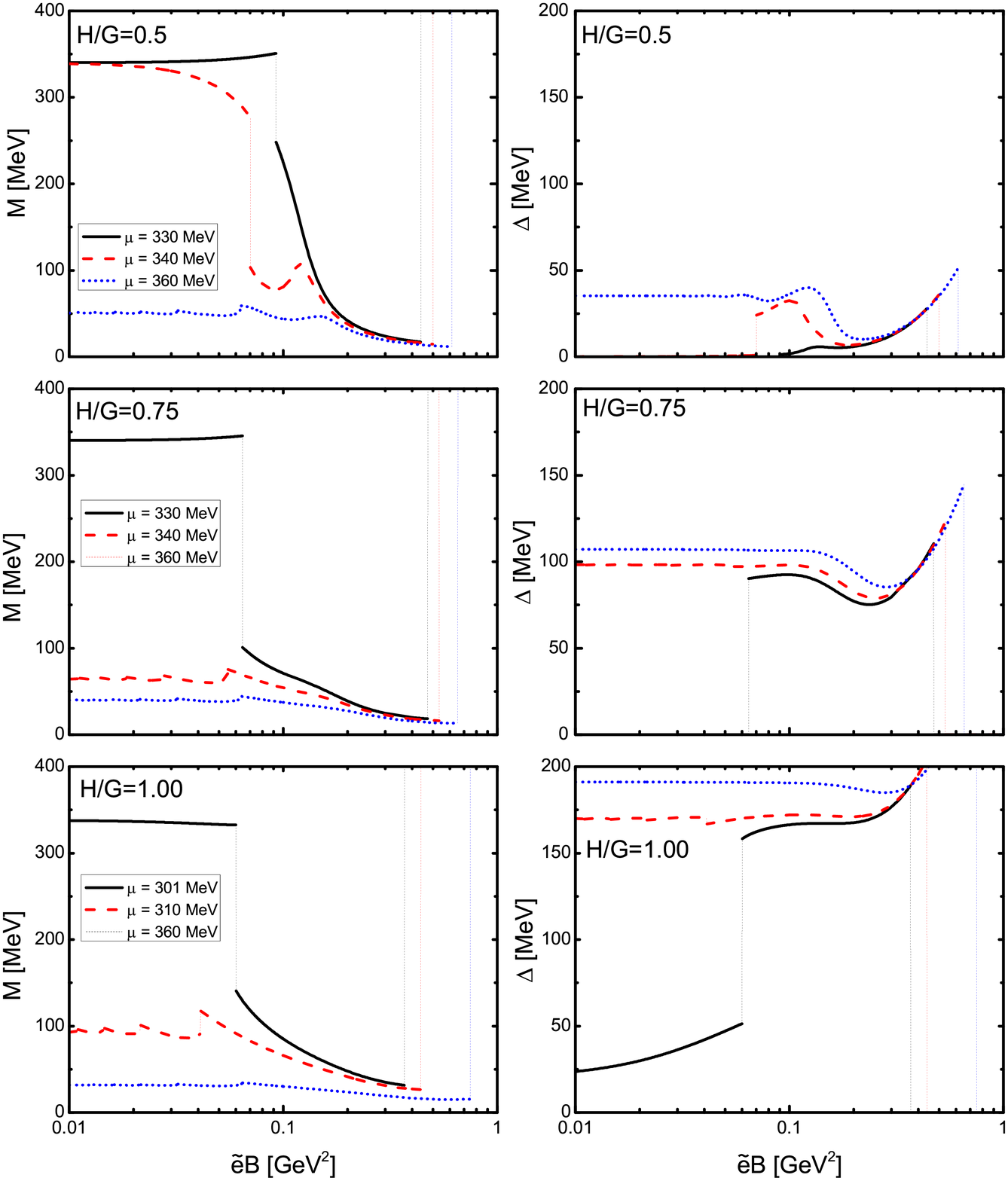}
\end{center}
\caption{M and $\Delta$ vs $\tilde eB$ within the MFIR scheme for
the three values of $H/G$ considered and several representative
values of $\mu$, set~1.} \label{fig:7}
\end{figure*}

In Fig.~\ref{fig:7}, we present the behavior of the order
parameters as a function of magnetic field, for all considered
values of $H/G$ and representative values of $\mu$. Once again, we
display the results for set 1 only, since no qualitatively
different behaviors arise for set 2. It was already seen in
Sec.III.A that in the $\mbox{B}$ phase, mass increases with
magnetic field and that $\Delta=0$. This result, which is a
manifestation of the magnetic catalysis effect, is exclusively
seen in vacuum. Since $\Delta=0$ in this phase, the behavior is
independent of $H/G$. When quark population is finite, several
possibilities arise depending on the magnetic field, parameter set
and coupling constant. For high chemical potentials
($\mu=360$~MeV), where chiral symmetry is restored for all values
of $H/G$, the mass exhibits a series of peak-like discontinuities.
They correspond to the already discussed vAdH transitions, where
the $k_{max}$ corresponding to the $|\tilde q|=1$ quark changes in
one unit. Due to the coupling to the mass, $\Delta$ also
oscillates. These oscillations become particularly large for
intermediate magnetic field, around $\tilde eB=0.1$~GeV$^2$. We
note that on increasing $H/G$, $M$ is shifted downwards and
$\Delta$ is shifted upwards. Also, all oscillations are smoothed
and become negligible compared to the scale of the order
parameters.

For $H/G=0.5$, $\mu=340$~MeV corresponds to the intermediate
$\mbox{C}$-type phase where quark population is finite and
$\Delta$ is small. We note that for $\tilde eB$ near zero, the
mass is close to the vacuum value, but gradually decreases as
$\tilde eB$ is increased. This feature is a manifestation of the
anticatalysis effect. On the other hand, $\Delta$ is very small
and increases only slightly, always remaining under $1$~MeV. The
jumps in both $M$ and $\Delta$ near $\tilde eB=0.1$~GeV$^2$,
correspond to a first order transition to an $\mbox{A}_1$ phase.
For larger magnetic fields, the mass decreases steadily until it
finally returns to the $\mbox{B}$ phase, which corresponds to the
high magnetic field branch of the main transition. In the lower
two panels of Fig.~\ref{fig:7}, we also see the behavior of $M$
and $\Delta$ in the mixed phase for $H/G=1$. Here, we can see that
the mass is almost independent of magnetic field in the $\mbox{D}$
phase (slightly decreasing), while the superconducting gap
increases from $25$~MeV to $50$~MeV. A few comments should be made
regarding the vAdH transitions. These transitions are a
characteristic of the chirally restored region of any NJL model
with magnetic field. In particular, in the model without diquark
pairing, there will be vAdH transitions for up quarks and down
quarks. In the rotated base, quarks with charges $|\tilde q|=1$
and $|\tilde q|=1/2$ couple to the magnetic field, of which only
the former yield first order transitions. However, there is a
vAdH-like cross over transition for $H/G=0.5$ and $0.75$ that
resembles a $|\tilde q|=1/2$ vAdH in the following sense. If we
were to set $H=0$, then $\Delta=0$, and the integrals in the
$|\tilde q|=1/2$ matter terms would transform to Heaviside
functions, cutting off the LL sum at a maximum k given by the
relation $\sqrt{\mu^2 - (M^2+k \tilde e B)}$ and therefore
producing vAdH transitions. If $H/G$ is made finite but small,
these discontinuities quickly disappear, but a remnant of these
transitions is still present because the associated quark number
susceptibility still exhibits peaks. However, they are smeared out
as $H/G$ is increased and for $H/G=0.5$ there is no trace left of
these transitions, except for the one corresponding to the LL
passage from $0$ to $1$. In set 1, we can see that this transition
is first order and has an end point after which it becomes a cross
over, as can be seen in the corresponding panel in
Fig.~\ref{fig:6}. This crossover is still present for $H/G=0.75$
and finally disappears for $H/G=1$. We make the observation that
since these are rather weak, we do not consider them to separate
distinct phases.

\section{Summary and Conclusions}

In the present work we explored the effects of magnetic field on
cold color superconducting quark matter in the framework of the
NJL-type model, using a regularization scheme in which the
contributions which are explicitly dependent on the magnetic field
turn out to be finite and, thus, do not require to be regularized.
Such a ``magnetic field independent regularization" (MFIR) scheme
can be considered an extension of the method described in e.g.
Ref.\cite{Menezes:2008qt} to the case in which color pairing
interactions are present. We compared the corresponding results
with those obtained through the regularization methods used in
previous
works\cite{Fukushima:2007fc,Noronha:2007wg,Fayazbakhsh:2010gc,Fayazbakhsh:2010bh,Mandal:2012fq}.
In those works, a regulator function was introduced in order to
separately render the contribution of each Landau level finite.
This, however, might lead to the appearance of non-physical
oscillations even when rather soft regulator functions are used,
while within the MFIR scheme these oscillations are completely
removed and, thus, results are easier to interpret. Note that in
the cases we investigated the smooth regulator case exhibits
barely no difference with respect to the MFIR scheme for $\tilde e
B$ up to $0.1$ GeV$^2$, and the above mentioned oscillations start
to become relevant from this value on. In respect of this, it
should be borne in mind that some estimates\cite{Ferrer:2010wz}
indicate that the magnetic fields at the center of magnetars can
be as large as $\tilde e B \sim 0.6$ GeV$^2$. The remaining of this
paper was devoted to investigate, for the MFIR regularization
scheme, the model parameter dependence of the behavior of
magnetized cold superconducting quark matter. We considered two
parameter sets that adjust to acceptable values of the dressed
masses, and that were already known to generate qualitatively
different phase diagrams in the non-superconducting case
\cite{Allen:2013lda}. Moreover, three representative values of the
coupling constant ratio $H/G$ were considered. We found that up to
$H/G=0.5$ superconducting effects are still small, and the phase
diagram remains basically unchanged. In particular, for set 1
there is an intermediate phase, in which quark population exists
but chiral symmetry is still strongly broken. The diquark gap in
this phase is finite but extremely small. It is connected to other
phases by first order transitions and a chiral crossover. As the
coupling ratio is increased, the two transitions surrounding this
phase move close to each other, causing it to disappear at
$H/G\sim0.65$. A relatively simple phase diagram hence exists for
a narrow range around the standard value $H/G=$0.75. If the
coupling ratio is increased beyond $0.94$, a mixed phase is
present in both parameter sets, where both condensates are large.
This phase already exists for zero magnetic field and we showed
that according to this model it extends in the magnetic field
direction for a relatively large range of $\tilde eB$,
disappearing for $\tilde eB \sim 0.1$GeV$^2$. It is bounded from
below by a second order transition. The nature of the vAdH
transitions was also discussed. We explained that in the rotated
base these exist only for the $|\tilde q|=1$ quark, and that even
though the $|\tilde q|=1/2$ quark couples to the magnetic field,
its coupling to the superconducting gap smears the vAdH
transitions out, leaving only a crossover remnant for the
transition between the two lowest $k$ values. Finally, we found
that the inverse catalysis phenomenon is observed in all the
cases, although less pronounced as $H/G$ increases.

Throughout this work we have concentrated on the analysis of the
impact of the novel regularization scheme on the model predictions
for the effects of magnetic field on symmetric two flavor cold
color superconducting quark matter. It is clear that, for
applications in the physics of compact stars, the neutrality and
$\beta$-equilibrium conditions should be taken into account.
Moreover, to address the effect on the description of the CFL-type
phases strangeness degrees of freedom have to be included. We
expect to report on these topics in forthcoming publications.

\section*{Appendix}

In this Appendix we present some details concerning the derivation
of Eq.(\ref{equa}). We start by considering Eq.(\ref{seis}). As a
first step towards the regularization of this expression we sum
and subtract the contribution for vanishing chemical potential. In
this way we get
\begin{equation}
P_{\left|\tilde{q}\right| = 1/2} = S_1 + S_2
\label{p12app}
\end{equation}
where
\begin{equation}
S_1 = \frac{\tilde eB}{\pi} \sum_{k=0}^\infty  \alpha_k
\int_{-\infty}^{\infty}\frac{dp_z}{2 \pi} \sqrt{p_z^2 + k \tilde
eB + M^2 + \Delta^2}.
\end{equation}
and
\begin{equation}
S_2 = \frac{\tilde eB}{2 \pi} \sum_{k=0}^\infty  \alpha_k \int_{-\infty}^{\infty}\frac{dp_z}{2 \pi} F(p_z^2 + k \tilde eB)
\end{equation}
where
\begin{equation}
F(z) = \sum_{s=\pm 1} \left[ \sqrt{ (\sqrt{z + M^2} + s \ \mu)^2 +
\Delta^2} - \sqrt{z + M^2 + \Delta^2}\right] \ .
\end{equation}
It is clear that both $S_1$ and $S_2$ are divergent. However $S_1$
has the standard form analyzed in Ref.\cite{Menezes:2008qt}, with $M^2
\rightarrow \Delta^2 + M^2$. Thus, following the steps described
in that reference we get
\begin{eqnarray}
S_1 &=& \frac{4}{\pi^2} \int_0^\Lambda\ dp\ p^2\ \sqrt{p^2 + M^2 +
\Delta^2} + \frac{(\tilde eB)^2}{2 \pi^2} \left[\xi'(-1,y) +
\frac{y-y^2}{2} \ln y + \frac{y^2}{4} \right] \label{t1}
\end{eqnarray} where $y = (M^2 + \Delta^2)/\tilde eB$ and,
as in Ref.\cite{Menezes:2008qt}, a 3D sharp cutoff has been
introduced to regularize the $\tilde eB$-independent contribution.

To regularize $S_2$ we add and subtract the corresponding
contribution in the absence of magnetic field. Namely,
\begin{equation}
S_2 = 4 \int \frac{d^3p}{(2 \pi)^3} F(p^2) + R
\label{t2}
\end{equation}
where
\begin{equation}
R = \frac{\tilde eB}{2 \pi} \sum_{k=0}^\infty  \alpha_k \int_{-\infty}^{\infty}\frac{dp_z}{2 \pi} F(p_z^2 + k \tilde eB)
- 4 \int \frac{d^3p}{(2 \pi)^3} F(p^2)
\label{rr}
\end{equation}
We prove in what follows that $R$ is finite. For this purpose we
start by rewriting its second term introducing cylindrical
coordinates. Then, once the angular integral is performed we get
\begin{equation}
\int \frac{d^3p}{(2 \pi)^3} F(p^2) = \int_{-\infty}^\infty \frac{d p_z}{2\pi} \int_0^\infty \frac{d\rho}{2\pi}\ \rho\ F(p_z^2+\rho^2) =
\frac{\tilde eB}{4\pi} \int_{-\infty}^\infty \frac{d p_z}{2\pi} \int_0^\infty dx \ F(p_z^2+ \tilde eB x)
\end{equation}
where the change of variables $\rho = \sqrt{\tilde e B x}$ has
been used in the last step. Replacing this expression in
Eq.(\ref{rr}) we get
\begin{equation}
R = \frac{\tilde eB}{2 \pi} \int_{-\infty}^{\infty}\frac{dp_z}{2 \pi} F(p_z^2) +
\frac{\tilde eB}{\pi} \int_{-\infty}^{\infty}\frac{dp_z}{2 \pi}
\left[ \sum_{k=1}^\infty F(p_z^2 + k \tilde eB) - \int_0^\infty dx\ F(p_z^2+ \tilde eB x) \right]
\label{rr2}
\end{equation}
For convenience we introduce at this stage $f(z) = F(\tilde eB
z)/\sqrt{\tilde eB}$ and perform the change of variables
$p=p_z/\sqrt{\tilde eB}$. Then we get
\begin{equation}
R = \frac{(\tilde eB)^2}{2 \pi^2} \int_0^{\infty} dp\ f(p^2) +
\frac{(\tilde eB)^2}{\pi^2} \int_0^{\infty} dp
\left[ \sum_{k=1}^\infty f(p^2 + k ) - \int_0^\infty\ dx\ f(p^2 + x) \right]
\label{rr3}
\end{equation}
where the fact that the integrands are even functions of $p$ has
been used. To proceed we notice that $f(z)$ is bounded in the
interval $0 \le z \le \infty$. Thus, to prove that the momentum
integrals are convergent it is enough to verify that the
corresponding integrands vanish sufficiently fast as
$p\rightarrow\infty$. For this purpose it is convenient to
consider the expansion of $f(z)$ for $z >> \mu^2/\tilde eB ;
(M^2+\Delta^2)/\tilde eB$. We get
\begin{equation}
f(z) = \sum_{m=1}^\infty \frac{C_m}{z^{m+ 1/2}}
\label{exp}
\end{equation}
where the values of $C_m$ for $m=1,2$ are
\begin{equation}
C_1 = \frac{\Delta^2 \mu^2}{(\tilde eB)^2} \qquad ; \qquad C_2 = \frac{\Delta^2 \mu^2}{(\tilde eB)^2}
\left[ \frac{\mu^2}{\tilde eB} - \frac{3}{2} \frac{M^2 + \Delta^2}{\tilde eB} \right]
\end{equation}
It is clear from Eq.(\ref{exp}) that for large $p$ the integrand
in the first term of $R$ goes as $1/p^{3}$ and, thus, the
corresponding integral is convergent. To prove the convergence of
the second term one can proceed as follows. Let us consider $N
>> \mu^2/\tilde eB ; (M^2+\Delta^2)/\tilde eB$ such that for $z >
N$ we can replace $f(z)$ by the expansion Eq.(\ref{exp}). As an
example, we note that for the range of values of $\mu$ and $\tilde
eB$ considered in this work, $\mu^2/\tilde eB$ and
$(M^2+\Delta^2)/\tilde eB$ are always smaller than $25$.
Accordingly, we have checked that if $N >> 25$ (e.g. N=250) the
use of the first few terms in the expansion Eq.(\ref{exp}) leads
to values which are in excellent agreement with those obtained
from the full expression. Then, denoting by $I$ the integrand in
the second term of Eq.(\ref{rr3}),  we have
\begin{eqnarray}
I &=& \left[ \sum_{k=1}^N f(p^2 + k ) - \int_0^N\ dx\ f(p^2 + x) \right] \nonumber\\
&+& \sum_{m=1}^\infty \ C_m
\left[ \sum_{k=N+1}^\infty \frac{1}{(p^2 + k)^{m+1/2}} - \int_N^\infty\  \frac{dx}{(p^2 + x)^{m+1/2}} \right]
\end{eqnarray}
Next we note that, since $m \ge 1$, the sum over $k$ in the second
term can be written in terms of the Hurwitz zeta function $\xi$
and that the integral over $x$ is convergent and can be explicitly
performed. Consequently we get
\begin{eqnarray}
I &=& \left[ \sum_{k=1}^N f(p^2 + k ) - \int_0^N\ dx\ f(p^2 + x) \right] \nonumber\\
&+& \sum_{m=1}^\infty \ C_m \left[ \xi\left(m+1/2, 1 + p^2 +N\right) - \frac{1}{m-1/2} \ \frac{1}{(p^2 + N)^{m-1/2}} \right]
\label{ieq}
\end{eqnarray}
Considering values of $p^2 >> N$ and making again use of
Eq.(\ref{exp}) it is easy to see that the two terms in the first
bracket of Eq.(\ref{ieq}) go as $N/p^3$ which implies that their
contributions to $R$ are finite. What remains is to show that the
leading contribution to the second bracket also goes faster than
$1/p$. This is not so obvious since for $m=1$ the second term does
go as $1/p$. Actually, as shown in what follows we need this term
to go like that in order to cancel a similar term arising from the
first term. In fact this is the main reason why the substraction
scheme proposed in this work was introduced. To show how this
cancellation occurs we consider the expansion of the Hurwitz zeta
function. For $x >> 1$ and $m \ge 1$ one has
\begin{equation}
\xi(m+1/2,1+x) = \frac{x^{-(m-1/2)}}{m-1/2}  - \frac{x^{-(m+1/2)}}{2}  + {\cal{O}}\left({x^{-(m+3/2)}}\right)
\end{equation}
Using this expansion it is easy to see that for $p^2 >> N$ we have
\begin{equation}
 \xi\left(m+1/2, 1 + p^2 +N\right) - \frac{1}{m-1/2} \ \frac{1}{(p^2 + N)^{m-1/2}}
\rightarrow - \frac{1}{2 p^{2m +1}} + {\cal{O}}\left({p^{-(2 m+3 )}}\right)
\end{equation}
This implies that the leading contribution comes from the $m=1$ term and it goes as $1/p^3$ ensuring the
convergence of the corresponding integral.

Having proved that $R$ is indeed finite only the first term in
$S_2$ requires regularization. Using a 3D sharp cutoff as before
we get
\begin{eqnarray}
S_2 &=& \frac{2}{\pi^2} \int_0^\Lambda\ dp\ p^2\ \left[ E_\Delta^+ + E_\Delta^- - 2
\sqrt{z + M^2 + \Delta^2}\right] \nonumber
\\
& & \qquad \qquad + \frac{(\tilde eB)^2}{2 \pi^2} \int_0^{\infty} dp
\left[ \sum_{k=1}^\infty \alpha_k \ f(p^2 + k ) - 2 \int_0^\infty\ dx\ f(p^2 + x) \right]
\label{s2reg}
\end{eqnarray}
where, as defined in the main text, $E_\Delta^\pm=\sqrt{ (\sqrt{p^2+ M^2} \pm \mu)^2 + \Delta^2}$.
Finally, replacing Eqs.(\ref{t1},\ref{s2reg}) in Eq.(\ref{p12app})
we immediately get Eq.(\ref{equa}).

%%%%%%%%%%%%%%%%%%%%%%%%%%%%%%%%%%%%%%%%%%%%%%%%%%%%%
%Figura 1
%%%%%%%%%%%%%%%%%%%%%%%%%%%%%%%%%%%%%%%%%%%%%%%%%%%%%%%

%%%%%%%%%%%%%%%%%%%%%%%%%%%%%%%%%%%%%%%%%%%%%%%%%%%%%
%Figura 2
%%%%%%%%%%%%%%%%%%%%%%%%%%%%%%%%%%%%%%%%%%%%%%%%%%%%%%%

%%%%%%%%%%%%%%%%%%%%%%%%%%%%%%%%%%%%%%%%%%%%%%%%%%%%%
%Figura 3
%%%%%%%%%%%%%%%%%%%%%%%%%%%%%%%%%%%%%%%%%%%%%%%%%%%%%%%

%%%%%%%%%%%%%%%%%%%%%%%%%%%%%%%%%%%%%%%%%%%%%%%%%%%%%
%Figura 4
%%%%%%%%%%%%%%%%%%%%%%%%%%%%%%%%%%%%%%%%%%%%%%%%%%%%%%%

%%%%%%%%%%%%%%%%%%%%%%%%%%%%%%%%%%%%%%%%%%%%%%%%%%%%%
%Figura 5
%%%%%%%%%%%%%%%%%%%%%%%%%%%%%%%%%%%%%%%%%%%%%%%%%%%%%%%

%%%%%%%%%%%%%%%%%%%%%%%%%%%%%%%%%%%%%%%%%%%%%%%%%%%%%
%Figura 6
%%%%%%%%%%%%%%%%%%%%%%%%%%%%%%%%%%%%%%%%%%%%%%%%%%%%%%%

%%%%%%%%%%%%%%%%%%%%%%%%%%%%%%%%%%%%%%%%%%%%%%%%%%%%%
%Figura 6A
%\begin{figure*}
%\includegraphics[scale=0.9]{HG05_set2.eps}
%\includegraphics[scale=0.9]{HG075_set2.eps}
%\includegraphics[scale=0.9]{HG1_set2.eps}
%\caption{Set 2, phase diagrams for different values of H/G}
%\label{figset340}
%\end{figure*}
%%%%%%%%%%%%%%%%%%%%%%%%%%%%%%%%%%%%%%%%%%%%%%%%%%%%%%%

%%%%%%%%%%%%%%%%%%%%%%%%%%%%%%%%%%%%%%%%%%%%%%%%%%%%%
%%Figura 7
%\begin{figure*}
%\begin{center}
%\includegraphics[scale=0.9]{FIG7a.eps}
%\includegraphics[scale=0.9]{FIG7b.eps}
%\end{center}
%\caption{M and $\Delta$ vs $\tilde{e}B$ for different ratios H/G and $\mu = 360~MeV$, set 1}
%\label{fig:6}
%\end{figure*}
%%%%%%%%%%%%%%%%%%%%%%%%%%%%%%%%%%%%%%%%%%%%%%%%%%%%%%%

%%%%%%%%%%%%%%%%%%%%%%%%%%%%%%%%%%%%%%%%%%%%%%%%%%%%%
%Figura 7
%%%%%%%%%%%%%%%%%%%%%%%%%%%%%%%%%%%%%%%%%%%%%%%%%%%%%%%

\end{document}